\documentstyle[aps,twocolumn]{revtex}
\begin{document}
\title{Observation of superconductivity in  Y$_2$PdGe$_3$, structurally same as  MgB$_2$}
\tighten
\baselineskip 0.5 cm
\author {E.V. Sampathkumaran$^{\dag}$ and Subham Majumdar}

\address{Tata Institute of Fundamental Research, Homi Bhabha Road, Colaba, Mumbai-400005, India}

\maketitle 

\begin{abstract}

The results of electrical resistance (1.4 - 300 K), magnetization (2-300 K) and heat-capacity (2 - 50 K) measurements in 
Y$_2$PdGe$_3$, found to crystallize in a AlB$_2$-derived hexagonal structure, are reported\footnote{The report of superconductivity in Y$_2$PdGe$_3$ has been communicated by us  to Physical Review B about few months back} . The results establish that this compound is  superconducting below 3 K. This obervation is interesting considering that this compound  is the first superconductor among the ternary members derived from the hexagonal AlB$_2$ structure. With superconductivity being {\it uncommon  even among binary alloys derived from  AlB$_2$ structure} and with recent excitement on the observation of high temperature superconductivity in Mg$B_2$, this finding gains importance.      

\end{abstract}
\vskip 0.5 cm
Recent observation\cite{1} of superconductivity with T$_c$ close to 40 K in MgB$_2$ has caused considerable excitement among condensed matter physicists.\cite{2,3} This compound crystallizes in the hexagonal AlB$_2$-type structure, which is made up of alternating of hexagonal layers of Al and graphite-like honeycomb layers of B atoms. This observation of superconductivity is further fascinating, particularly when one notes that this structure is generally unfavorable for superconductivity. Thus, for instance,  among the rare-earth disilides and germanides having the same electron concentration (YSi$_2$, YGe$_2$, LaSi$_2$ and LaGe$_2$), only those with  the $\alpha$-ThSi$_2$ structure (YGe$_2$, LaSi$_2$ and LaGe$_2$, structurally slightly different from that of AlB$_2$) have been found to be superconducting;\cite{4,5,6} even for ThSi$_2$, the $\alpha$-tetragonal form exhibits a higher T$_c$ (3.16 K) than the $\beta$-form which is of AlB$_2$-type (2.4 K).\cite{7,8} Thus, it is of interest to search for superconductors among compounds crystallizing in the AlB$_2$-type structure, particularly to look for the ones not containing B in view of a proposal that B, being a light atom, plays a vital role to favor superconductivity in this structure.

During last few years, we have been actively involved in synthesizing new compounds of the type R$_2$XY$_3$, with crystal structures derived from that of AlB$_2$ (Ref. 9). Though we noticed several interesting magnetic anomalies,\cite{10,11,12,13,14,15,16,17,18,19,20} in particular relevant to the fields of Kondo lattices and heavy-fermions, non-Fermi-liquids, colossal magnetoresistance, magnetic refrigeration and possibly low-dimensional magnetism, it is to be noted that none of the present family of compounds has been found (or earlier reported) to be superconducting. Here, we report the formation of the compound, Y$_2$PdGe$_3$,  in the AlB$_2$-derived structure and find it to be the first superconductor among this family of ternary compounds (that is, with 2:1:3 stoichiometry) with a superconducting transition temperature (T$_c$) of 3 K.

The samples were prepared by arc melting stoichiometric
amounts of constituent elements in an inert atmosphere.  The ingots were homogenised
in an evacuated, sealed quartz tube at 850 C. X-ray diffraction patterns were 
obtained employing Cu K$_\alpha$ radiation and  the patterns confirm that the
compounds  crystallize in an AlB$_2$-derived hexagonal
structure.  There is an additional weak line (about 5\%), which we believe is due to a  parasitic phase with the stoichiometry 1:2:2. We do not find any superstructure line in the X-ray diffraction patterns within the detection limit and hence we believe that there is no doubling of unit-cell parameters (a= 4.192 \AA and c= 4.000 \AA) unlike in some other isostructural compounds.\cite{1,20} The electrical resistivity ($\rho$) measurements (T= 1.4-300 K) were performed by a conventional four-probe method employing a silver paint for making electrical contacts.  The field-cooled (FC) and the zero-field-cooled (ZFC) magnetic susceptibility ($\chi$) behavior (H= 25 Oe)  below 10 K,  and the heat-capacity (C) data (2.5-30 K) by semi-adiabatic heat-pulse method were also obtained.

With respect to the high temperature $\rho$ behavior, there is a normal metallic behavior of $\rho$ with the variation of T. However, at 3 K, there is a sharp drop of $\rho$ (Fig. 1) to zero (within the accuracy of nanovoltmeter employed to measure the voltage drop across the leads) as T is lowered, as if this compound is superconducting below (T$_c$=) 3 K. To support this finding, the $\chi$ has also been measured at low fields (25 Oe) and we find that there is an onset of strong diamagnetism below 3 K and the divergence of ZFC and FC $\chi$ is typical of that expected for type-II superconductors. A comparison of the magnitude of the value of ZFC-$\chi$ at 2 K with  that of other known standard superconductors like Pb establishes bulk nature of superconductivity. The plot of M versus H (see Fig. 1, inset) at 1.7 K is typical of that of type-II superconductors with a value of lower critical field of about 400 Oe. In order to render further support to the bulk nature of superconductivity, C was also measured down to 2.5 K and the upturn observed in the data obtained below 3 K (see Fig. 1, bottom inset) is sufficient to establish bulk superconductivity; the broadened signal  around the transition could possibly due to inhomogeneities. The values of the Debye temperature and the electronic term ($\gamma$) inferred from the data in the range 5 - 15 K turn to be about 130 K and 2.5mJ/mol K$^2$ respectively. If one employs this value of $\gamma$, the value of $\Delta$C/$\gamma$T$_c$ turns to be close to 3, which is far away from the weak coupling value of 1.35.

To conclude, we report the formation of a new compound, Y$_2$PdGe$_3$, in the hexagonal AlB$_2$-structure and studied its physical characteristics. This compound  has been identified to be the lone superconductor till date in this family of ternary rare-earth compounds. We hope that the observation of superconductivity in a AlB$_2$-type compound, not containing B, may be helpful for overall theoretical understanding of high T$_c$ in MgB$_2$.Finally we would like to state that a weak superconducting signal around 7 K has been obtained even in the composition Y$_2$PdSi$_3$\cite{21}. 
\vskip 0.3 cm  

\noindent $\dag$ {\it E-mail address: sampath@tifr.res.in

\begin{figure}
\caption{Electrical resistivity and magnetic susceptibility (H= 25 Oe) behavior at low temperatures for Y$_2$PdGe$_3$ to highlight the features due to superconducting transition. The top inset shows the field dependence of M at 1.7 K for increasing and then decreasing field for the ZFC state of the specimen and the bottom inset shows the plot of heat capacity versus temperature to highlight the feature around superconducting transition. The line through the $\rho$ data serves as a guide to the eyes.} 
\end{figure}

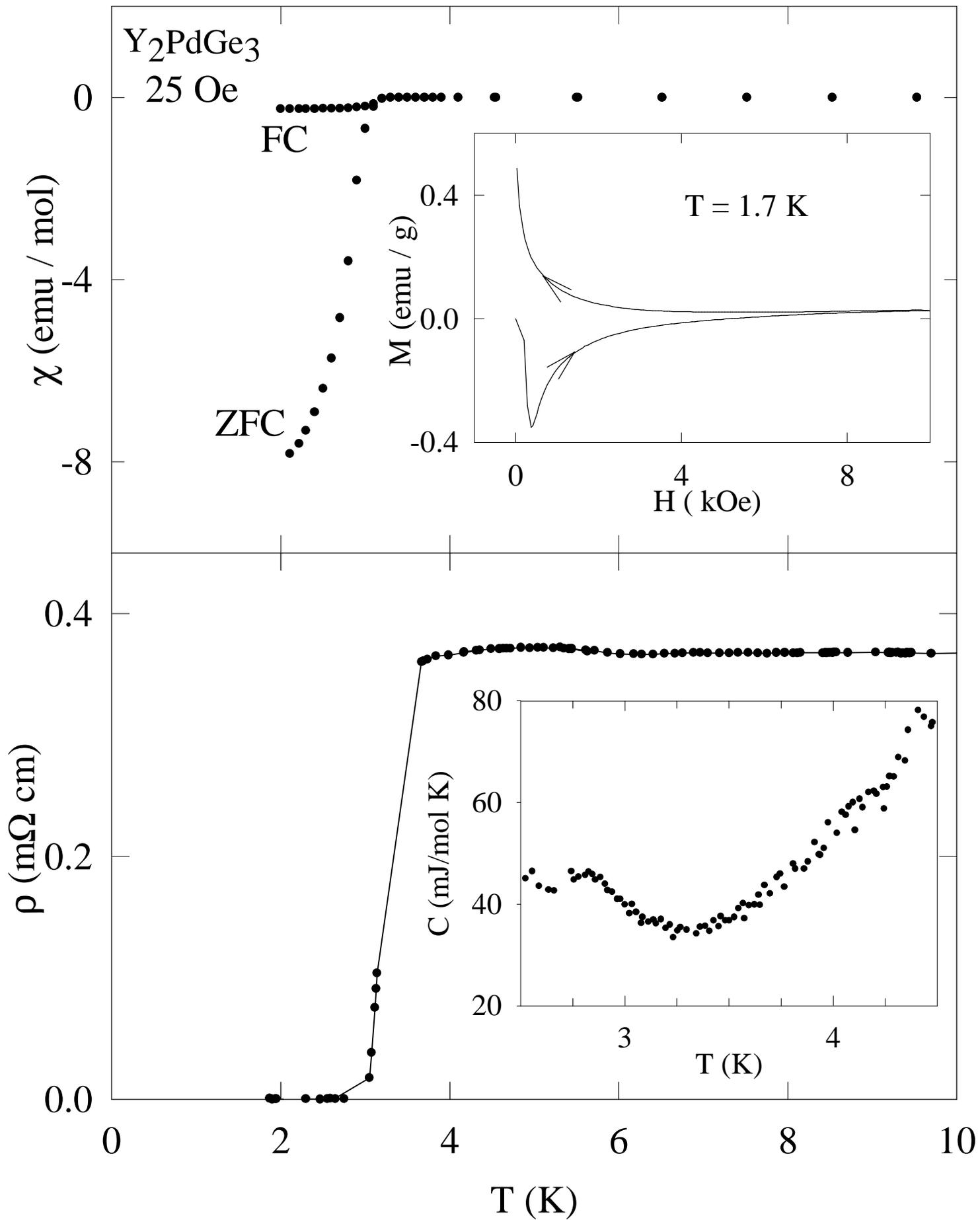
\end{document}